\documentclass[a4paper, oneside, twocolumn, notitlepage, 10pt]{extarticle_ecoc}
\usepackage{ecoc}

\usepackage{bm}

\usepackage{mleftright}

\usepackage{subcaption}

\def\VAEEonlyCPR{\textit{VAEE+CPR}}
\def\VAEEtrainCPR{\textit{PC-VAEE}}      %

\definecolor{ColCM}{rgb}{0,0.4470,0.7410}
\definecolor{ColVAEE}{rgb}{0.8500,0.3250,0.0980}
\definecolor{ColVAEEonlyCPR}{rgb}{0.4940,0.1840,0.5560}
\definecolor{ColVAEEtrainCPR}{rgb}{0.4660,0.6740,0.1880}

\definecolor{KITgreen}{rgb}{0,.59,.51}
\definecolor{KITblue}{rgb}{.27,.39,.66}
\definecolor{KITred}{rgb}{.63,.13,.13}
\definecolor{KITorange}{rgb}{.87,.60,.10}
\definecolor{KITyellow}{rgb}{.98,.89,0}
\definecolor{KITpalegreen}{RGB}{130,190,60}
\definecolor{KITpurple}{RGB}{160,0,120}
\definecolor{KITcyanblue}{RGB}{80,170,230}

\newcommand{\Meq}{M_{\mathrm{eq}}}
\newcommand{\Mest}{M_{\mathrm{est}}}

\newcommand{\Ncpr}{N_{\mathrm{cpr}}}

\newcommand{\Nb}{N_{\mathrm{batch}}}	
\newcommand{\taupmd}{\tau_{\mathrm{pmd}}}
\newcommand{\Lcd}{L_{\mathrm{cd}}}

\newcommand{\mlr}[1]{\mleft(#1\mright)}

\DeclareMathOperator{\mean}{mean}

\def\j{{\mathrm{j}}}
\def\e{{\mathrm{e}}}

\usepackage[nohyperlinks,nolist]{acronym}
\acrodef{AWGN}[AWGN]{additive white Gaussian noise}

\acrodef{BMI}[BMI]{binary mutual information}
\acrodef{BPS}[BPS]{blind phase search}

\acrodef{CD}[CD]{chromatic dispersion}
\acrodef{CM}[CM]{constant-modulus}
\acrodef{CMA}[CMA]{constant-modulus algorithm}
\acrodef{CPR}[CPR]{carrier-phase recovery}

\acrodef{DSP}[DSP]{digital signal processing}

\acrodef{ELBO}[ELBO]{evidence lower bound}
\acrodef{ECL}[ECL]{external-cavity laser}

\acrodef{FEC}[FEC]{forward error correction}
\acrodef{FIR}[FIR]{finite impulse response}

\acrodef{JCAS}[JCAS]{joint communication and sensing}

\acrodef{MIMO}[MIMO]{multiple-input multiple-output}
\acrodef{ML}[ML]{maximum likelihood}
\acrodef{MSE}[MSE]{mean squared error}
\acrodef{MCF}[MCF]{multi-core fiber}
\acrodef{MD}[MD]{modal dispersion}

\acrodef{PCS}[PCS]{probabilistic constellation shaping}
\acrodef{PMD}[PMD]{polarization mode dispersion}

\acrodef{QAM}[QAM]{quadrature amplitude modulation}

\acrodef{RRC}[RRC]{root-raised cosine}
\acrodef{RC-4CF}[RC-4CF]{randomly-coupled 4-core fiber}

\acrodef{SER}[SER]{symbol error rate}
\acrodef{SSMF}[SSMF]{standard single mode fiber}
\acrodef{SNR}[SNR]{signal-to-noise ratio}
\acrodef{sps}[sps]{samples per symbol}
\acrodef{SDM}[SDM]{space-division multiplexing}

\acrodef{VAEE}[VAEE]{variational-autoencoder-based equalizer}

\begin{document}
\selectlanguage{english}    %

\title{Novel Phase-Noise-Tolerant Variational-Autoencoder-Based Equalization Suitable for Space-Division-Multiplexed Transmission}%

\author{
    Vincent Lauinger\textsuperscript{(1),$\ast$}, Lennart Schmitz\textsuperscript{(2)}, Patrick Matalla\textsuperscript{(2)}, 
    Andrej Rode\textsuperscript{(1)}, 
    Sebastian Randel\textsuperscript{(2)},\\ Laurent Schmalen\textsuperscript{(1)}
}

\maketitle                  %

\begin{strip}
    \begin{author_descr}

        \textsuperscript{(1)} Communications Engineering Lab (CEL), Karlsruhe Institute of Technology (KIT)

        \textsuperscript{(2)} Institute of Photonics and Quantum Electronics (IPQ), Karlsruhe Institute of Technology (KIT)

        \textsuperscript{$\ast$} Corresponding author:
        \textcolor{blue}{\uline{vincent.lauinger@kit.edu}}

    \end{author_descr}
\end{strip}

\renewcommand\footnotemark{}
\renewcommand\footnoterule{}

\begin{strip}
    \begin{ecoc_abstract}
        We demonstrate the effectiveness of 
        a novel phase-noise-tolerant, variational-autoencoder-based equalization scheme for space-division-multiplexed (SDM) transmission in an experiment over 150km of randomly-coupled multi-core fibers. \textcopyright2025 The Author(s)
    \end{ecoc_abstract}
\end{strip}

\section{Introduction}
\Ac{SDM} allows to overcome the information capacity limitations of \acp{SSMF} by exploiting multiple spatial channels~\cite{Richardson_nature} and, potentially, decrease the energy consumption per bit by sharing hardware. %
While randomly-coupled \acp{MCF} are beneficial, e.g., in their tolerance to nonlinearities, they require more challenging \ac{MIMO} \ac{DSP}~\cite{hayashi_RCMCF}. By avoiding the overhead from pilot transmission, blind algorithms can increase the net data rate or the robustness of the system by allowing a higher \ac{FEC} overhead. The state-of-the-art blind \ac{CMA} is sub-optimal for higher-order modulation formats and suffers from convergence issues for \ac{PCS}~\cite{zervas1991effects}, which became integral to optical communication systems due to its benefits, e.g., simple rate adaption~\cite{BuchaliJLT}. A promising alternative is the \ac{VAEE}~\cite{caciularu2018blind} which got attention in the past years~\cite{lauinger2022blind,tomczyk_VAE,rode_oft2025,burshtein_semisupervised,lauinger_bootstrap,bauer_wsa24}, and also enables \ac{JCAS} by combining \ac{ML}-approximating channel estimation with excellent equalization performance. However, to the best of our knowledge, this approach has not been analyzed with data from optical transmission experiments. %

\section{Variational-Autoencoder-Based Equalization}
The \ac{VAEE} leverages statistical inference techniques %
to approximate the \ac{ML} channel estimate and, as a byproduct, equalizes the received signal. The concept boils down to a loss function $\mathcal{L}$, often denoted as the \ac{ELBO}, which only depends on the statistics of the transmitted signal, i.e., the prior distribution, %
when using, e.g., a butterfly structure with \ac{FIR} filters as channel estimator~\cite{lauinger2022blind}. Hence, this approach is blind in the sense that it does not require pilot tones or a known transmit sequence. 
\begin{figure}[t!]
    \centering
	\includegraphics{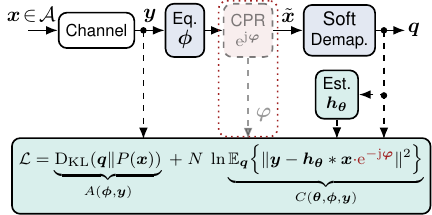} 
	\caption{Block diagram of the \ac{VAEE} with equalizer block (Eq.), channel estimator (Est.) and soft demapper (Soft Demap.). The \textcolor{KITred}{red} elements, i.e., the pale box including the \ac{CPR} block and the phase rotation in the loss function, are only applicable for the proposed \VAEEtrainCPR.} \label{Fig:VAE_scheme_both}
\end{figure}
The basic scheme of the \ac{VAEE} is depicted in Fig.~\ref{Fig:VAE_scheme_both}. The noisy received signal $\bm{y}$ is fed into an arbitrary equalizer function, which is parameterized by weights $\bm{\phi}$, and followed by a soft demapper to obtain estimates $\bm{q}$ of the posterior distribution. These soft values can be used for soft \ac{FEC}~\cite{haizheng_SC_LDPC} and are input to the loss function $\mathcal{L}$, which consists of two terms: $A$ is the relative entropy (also known as Kullback-Leibler divergence) %
between the real prior and the approximated posterior $\bm{q}$, which favors output distributions similar to the prior; and $C$ is the expectation $\mathbb{E}_{\bm{q}}$ (with respect to $\bm{q}$) of the \ac{MSE} between the noisy received signal $\bm{y}$ and the projected received signal, which can be computed from the estimated channel taps $\bm{h}_\theta$, the estimated posterior $\bm{q}$ and the discrete modulation alphabet~\cite{lauinger2022blind}. %
An arbitrary optimizer, e.g., Adam~\cite{kingma2014adam} or simple gradient descent, can be used to minimize the loss by jointly updating the equalizer and channel estimator weights. 

\begin{figure*}[t]
    \centering
    \includegraphics{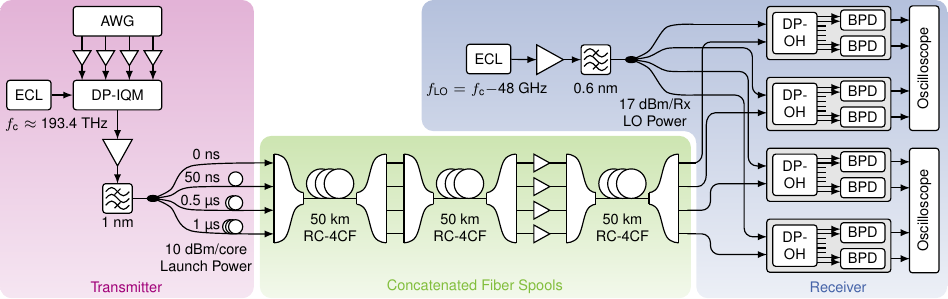}	
    \caption{Block diagram of the experimental SDM setup employing a randomly-coupled 4-core fiber~\cite{Matalla_JLT_SDM}.}
    \label{fig:exp_setup}
\end{figure*}
\begin{figure*}[b!]
\vspace*{-2ex}
    \captionsetup[subfigure]{labelformat=empty}
    \begin{subfigure}{.5\textwidth}
        \centering
        \includegraphics{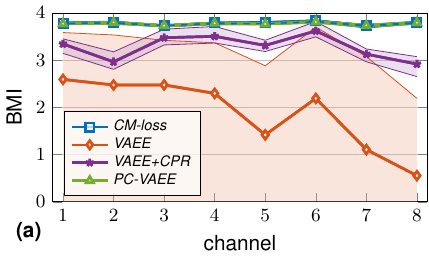} 
    \end{subfigure}
    \begin{subfigure}{.5\textwidth}
        \centering
        \includegraphics{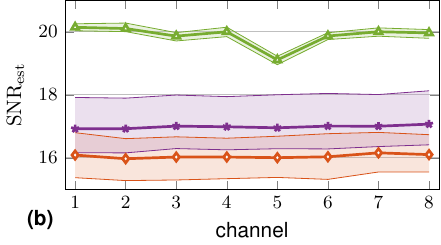} 
    \end{subfigure}
    \caption{Results for the \ac{SDM} transmission experiment with the \acsp{RC-4CF}: bit-wise mutual information (BMI) in (a) and estimated signal-to-noise ratio in (b). The markers denote the $\mean$, the shaded area is limited by the corresponding $\min$ and $\max$ values.} \label{Results:BMI_real_data}
\end{figure*}

An important aspect of the \ac{VAEE} is its inherent phase-sensitive nature, which is a major difference to the popular \ac{CM} loss~\cite{godard}. 
As a consequence, the \ac{VAEE} is able to track and compensate phase fluctuations to some degree; however, strong phase noise requires perpetual adaptation and, eventually, may prevent proper convergence of the equalizer. In the latter case, a trailing \ac{CPR} algorithm can be introduced after the \ac{VAEE} by leaving the structure as introduced before (where $\bm{q}$ is used to compute the loss), while compensating the remaining phase error of the equalized signal $\tilde{\bm{x}}$ with a \ac{CPR} and getting a phase-corrected estimate $\tilde{\bm{q}}\neq\bm{q}$ for further evaluation. Similarly to \ac{CM}-based equalizers, the \ac{CPR} is not part of the update procedure and, thus, does not have to be differentiable. However, potential convergence issues due to perpetual adaptation remain. In the following, this variant of the \ac{VAEE} is denoted \VAEEonlyCPR.

We propose a non-trivial approach to include \ac{CPR} to the update procedure as indicated by the red elements in Fig.~\ref{Fig:VAE_scheme_both}, where two points have to be considered: first, the \ac{CPR} has to be differentiable, which is, e.g., not the case for the plain \ac{BPS} algorithm~\cite{pfau2009hardware}. However, we can substitute the non-differentiable $\mathrm{argmax}$ operation with a $\mathrm{softmax+temperature}$ during gradient propagation as used for end-to-end learning~\cite{rodeOFC}. Second, the loss function has to be adapted since the compensated phase error in the equalized signal has to be learned by the channel estimator as well~\cite{Randel11}. As indicated by the red term in the loss function, this problem can be solved by rotating the projected transmit symbols with the negative phase offset estimate $-\bm{\varphi}$. In the following, this variant of the \ac{VAEE} is denoted \VAEEtrainCPR.

As a reference, we update the equalizer, consisting of the same butterfly structure with \ac{FIR} filters as the \ac{VAEE} variants, with an Adam optimizer using the \ac{CM}-loss~\cite{godard}. %
To implement the whole setup, we use \textit{MOKka}~\cite{mokka}, an open-source library for machine learning in communications, and \textit{Weights\&Biases}~\cite{wandb} to find the best-fitting hyperparameters for each algorithm.

\section{Experimental Setup and Results}
We validate the \ac{VAEE} in an \ac{SDM} transmission experiment, where eight \qty{90}{GBd} 16-QAM signals are transmitted through \qty{150}{km} of \ac{RC-4CF}~\cite{hayashi2017}, as thoroughly described in~\cite{Matalla_JLT_SDM} and depicted in Fig.~\ref{fig:exp_setup}. %
The \ac{ECL} has a nominal linewidth of \qty{100}{kHz} and the fiber's overall group delay spread caused by \ac{MD} is specified to \qty{10}{}$-$\qty{12}{ps\per\sqrt{km}}. %
At the receiver, the \ac{DSP} chain includes resampling to two \ac{sps}, carrier-frequency offset compensation based on a single polarization~\cite{menno2024}, \ac{CD} compensation using the overlap-and-save algorithm~\cite{xu2011}, and a joint clock recovery~\cite{Matalla_JLT_SDM}.

The equalizer consists of a complex-valued $8\!\times\!8$ butterfly structure of \ac{FIR} filters with $\Meq\!=\!151$~taps, which are updated batch-wise over $\Nb$~symbols. The batches overlap such that no extra padding is required for the filters. %
The channel estimators of the \ac{VAEE} are constructed similarly to the equalizers. If \ac{CPR} is applicable, we use a differentiable \ac{BPS} algorithm with temperature $T\!=\!0.01$, 40~test angles per quadrant, and triangular averaging window of length~$\Ncpr$.
After each frame of \qty{10000}{} %
symbols, we estimate the \ac{BMI} per channel. For each run, 100~frames are processed and the last 20~estimates are stored. Thus, each depicted value is the $\{\min,\ \max,\ \mathrm{mean}\}$ over 200~estimates from 10~independent runs of the same configuration. 
\begin{figure*}[tb!]
   \captionsetup[subfigure]{labelformat=empty}
   \begin{subfigure}{.5\textwidth}
   		\centering
   		\includegraphics{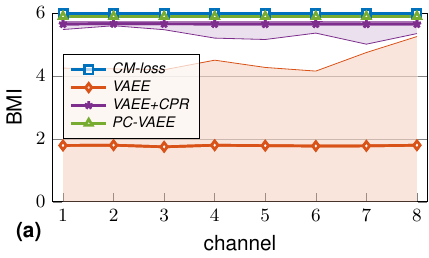}
   \end{subfigure}
   \begin{subfigure}{.5\textwidth}
   		\centering
   		\includegraphics{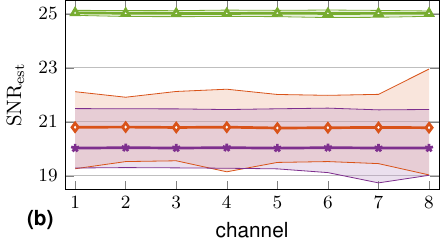}
	\end{subfigure}
    \vspace*{-2ex}
    \caption{Results for the simulation with uniform 64-QAM at \qty{90}{GBd} and very high linewidth $\delta_f=\qty{1}{MHz}$: bit-wise mutual information (BMI) in (a) and estimated signal-to-noise ratio in (b). The markers denote the $\mean$, the shaded area is limited by the corresponding $\min$ and $\max$ values.
    } \label{Results:BMI_simu_data}
\end{figure*}
A pilot-aided \ac{MSE} loss with integrated supervised phase recovery is used for pre-convergence over one frame, i.e., \qty{10000}{} symbols. Note that the resulting $\mathrm{BMI}\!<\!0.05$ (\ac{SER}$>\!0.9$) is not sufficient to switch to a decision-directed \ac{MSE} loss. 

As shown in Fig.~\ref{Results:BMI_real_data}(a), the reference blind \textit{CM-loss} as well as the proposed \VAEEtrainCPR \ converge properly while the \VAEEonlyCPR \ and the plain \textit{VAEE} suffer from penalties. For the latter two, the $\max$ of the \ac{BMI} is significantly higher than the $\mathrm{mean}$, indicating that the filters converged but the rapidly varying phase offset consistently prevents proper demapping. The difference in the shaded areas, i.e., the area between the corresponding $\min$ and $\max$ \ac{BMI}, of the \textit{VAEE} and the \VAEEonlyCPR \ is only due to compensation of the remaining phase offset, since the update procedure is the same. Thus, only the proposed \VAEEtrainCPR \ fully benefits from \ac{CPR}. 
As the \ac{VAEE} acts as \ac{ML}-approximating channel estimators~\cite{lauinger2022blind}, it enables \ac{JCAS} by monitoring the \ac{SNR} and impulse responses of the channel. Figure~\ref{Results:BMI_real_data}(b) depicts the estimated channel $\mathrm{SNR}_\mathrm{est}$ corresponding to the processing of Fig.~\ref{Results:BMI_real_data}(a). 
If the algorithms have not converged properly (e.g., the \textit{VAEE} and the \VAEEonlyCPR), the $\mathrm{SNR}_\mathrm{est} = P_\mathrm{sig}\cdot N/C$ is underestimated due to its inverse dependence on the $C$-term of $\mathcal{L}$~\cite{lauinger2022blind}. %

\section{Simulation Setup and Results}
To specifically analyze the influence of phase noise to the proposed \acp{VAEE}, %
we simulate the transmission of uniform 64-\ac{QAM} over a simple model of an uncoupled 4-core \ac{MCF} at $R_\mathrm{S}\!=\!\qty{90}{GBd}$ and different laser linewidth $\delta_f$. The transmitted pulses are shaped by a \ac{RRC} filter with roll-off of 0.2 and the received sequence is sampled with 2~\ac{sps}. Each core is modeled as linear optical dual-polarization channel with the frequency domain channel matrix~\cite{lauinger_bootstrap} 
\begin{align*}
	\bm{H}\mlr{f} &= \bm{R}\mlr{\gamma_{\text{hv}}} \begin{pmatrix} \e^{\j\pi \tau_{\text{pmd}}  f } & 0 \\ 0 & \e^{-\j\pi \tau_{\text{pmd}}  f } \end{pmatrix}  \e^{-\j2\pi^2 \beta_{\text{cd}} L_{\text{cd}}  f^2 }  %
\end{align*}
with rotation matrix $\bm{R}$ describing a rotation by $\gamma_{\text{hv}}\!=\!\frac{\pi}{10}$, differential group delay $\taupmd\! = \!\sqrt{\qty{1000}{}}\cdot\qty{0.1}{ps}$ causing first-order \ac{PMD}, and residual \ac{CD} of $\beta_{\text{cd}}\Lcd\!=\!-\qty{26}{\pico\second^2} \cdot \qty{2}{}$~\cite{agrawal2010fiber}. %
The resulting eight signals are randomly permuted and distorted by different phase noise realizations of a Wiener process with variance $\sigma^2_{\Delta\varphi}\!=\!2\pi \delta_f/R_\mathrm{S}$ as well as \ac{AWGN} such that the \ac{SNR} is \qty{25}{dB}. %
We apply the same averaging scheme to the estimates and construct the equalizer, channel estimator and \ac{CPR} similarly to the section above with $\Meq\!=\!\Mest\!=\!51$ taps. 

The results are depicted in Fig.~\ref{Results:BMI_simu_data}. Similarly to the experimental results, the \VAEEtrainCPR \ converges reliably for uniform 64-QAM even with very high linewith ($\delta_f\!=\!\qty{1}{MHz}$), as depicted in Fig.~\ref{Results:BMI_simu_data}(a)~and~(b), and estimates the simulated \ac{SNR} of \qty{25}{dB} very accurately. %
Interestingly, the \textit{VAEE} and \VAEEonlyCPR, which have the same update procedure, only show relatively small penalties in their $\max$ \ac{BMI} for the fast-varying phase distortions. 
While the trailing \ac{CPR} compensates remaining phase errors for the \VAEEonlyCPR \, the performance of the \textit{VAEE} is degraded significantly. 
The relatively worse \ac{SNR} estimation reveals that, in this scenario, the algorithms rely more on the $A$-term of $\mathcal{L}$ (KL-divergence) than the $C$-term, which is relevant for \ac{SNR} estimation~\cite{lauinger2022blind}.

\section{Conclusion}
We demonstrated the potential of \acp{VAEE}, in particular, the proposed \VAEEtrainCPR, in \ac{SDM} setups, especially in combination with \ac{JCAS}. A good \ac{CPR} is crucial for application in fiber-optic communication systems, where phase noise can be significant. Future work may focus on systems which employ \ac{PCS}, the \ac{JCAS} aspect, or the further applications. %

\clearpage
\section{Acknowledgements}
This work has received funding from the German Federal Ministry of Education and Research (BMBF) under grant agreement 16KIS1420 (STARFALL) and the European Research Council (ERC) under the EU’s Horizon 2020 research and innovation programme (grant agreement 101001899).

\defbibnote{myprenote}{%
Citations must be easy and quick to find. More precisely:
\begin{itemize}
    \item Please list all the authors. 
    \item The title must be given in full length. 
    \item Journal and conference names should not be abbreviated but rather given in full length.
    \item The DOI number should be added incl. a link.
\end{itemize}
}
\printbibliography

\vspace{-4mm}

\end{document}